\documentclass[12pt,epsf]{article}
\usepackage[xdvi]{graphicx}
\newcommand{\beq}{\begin{equation}}
\newcommand{\eeq}{\end{equation}}
\newcommand{\bea}{\begin{eqnarray}}
\newcommand{\eea}{\end{eqnarray}}

\def\pe2{p_E^2}

\begin{document}
\newcommand{\mpl}{M_{\mathrm{Pl}}}
\setlength{\baselineskip}{18pt}
\begin{titlepage}
\begin{flushright}
OU-HET 587/2007 \\
KOBE-TH-07-08
\end{flushright}

\vspace{1.0cm}
\begin{center}
{\Large\bf A Simple Model of Direct Gauge Mediation \\
\vspace*{5mm}
of Metastable Supersymmetry Breaking} 
\end{center}
\vspace{20mm}

\begin{center}
{\large
Naoyuki Haba$^*$\footnote{e-mail: haba@het.phys.sci.osaka-u.ac.jp}
and Nobuhito Maru$^\dag$\footnote{e-mail: 
maru@people.kobe-u.ac.jp}}
\end{center}
\vspace{1cm}
\begin{center}
$^*${\it Department of Physics, Osaka University,
Toyonaka, Osaka 560-0043, Japan}\\
\vspace*{5mm}
$^\dag${\it Department of Physics, Kobe University,
Kobe 657-8501, Japan}\\
\end{center}
%
%   Abstract
%
\vspace{15mm}
\centerline{\large\bf Abstract}
\vspace*{1cm}
We construct a model of direct gauge mediation of metastable SUSY breaking 
by simply deforming the Intriligator, Seiberg and Shih model 
in terms of a dual meson superpotential mass term. 
No extra matter field is introduced. 
The deformation explicitly breaks a $U(1)_R$ symmetry 
and a pseudo moduli have a nonzero VEV at one-loop. 
Our metastable SUSY breaking vacuum turns out to be sufficiently long-lived. 
By gauging a subgroup of flavor symmetry, our model can directly couple to 
the standard model, which leads to nonvanishing gaugino mass generation. 
It is also shown that our model can evade the Landau pole problem. 
We show the parameters in the SUSY breaking sector are 
phenomenologically constrained. 
\end{titlepage}

%
% Body
%
%%%%%%%%%%%%%%%%%%%%%%
\section{Introduction} 
%%%%%%%%%%%%%%%%%%%%%% 
Dynamical supersymmetry (SUSY) breaking is a convincing scenario 
to solve the gauge hierarchy problem \cite{Witten1}. 
According to Witten index \cite{Witten2}, 
the models of dynamical SUSY breaking were 
limited to the chiral models since the vector-like models 
except for the models discussed in \cite{IY, IT} do not break SUSY 
similar to super Yang-Mills theories. 
Building such chiral models with dynamical SUSY breaking was very troublesome. 
Once one can obtain SUSY breaking models, 
the next issue is how to transmit the SUSY breaking to our real world. 
The gauge mediation where the SUSY breaking effects are transmitted 
to our world by the Standard Model(SM) gauge interactions 
is one of the attractive framework 
since the gauge interaction is flavor-blind, 
thus there is no SUSY FCNC problem \cite{GM, DNS}. 
If we try to communicate the SUSY breaking directly to our world 
by gauging a subgroup of the flavor symmetry 
in the chiral SUSY breaking models known before 
and identifying it with the SM gauge group, 
it is likely that the QCD coupling blows up 
below the GUT scale \cite{ADS}. 
This is because the number of color is always 
larger than the flavor number in such chiral SUSY breaking models, 
which leads to the large number of messengers. 
This led to the introduction of the messenger sector \cite{DNS}, 
but the sector was in general quite complicated. 
Some models of direct gauge mediation 
avoiding the Landau pole problem were constructed so far \cite{PT}-\cite{Maru}.

Recently, Intriligator, Seiberg and Shih (ISS) have discovered 
a metastable SUSY breaking vacuum in ${\cal N}=1$ massive (but light) SUSY QCD 
in a free magnetic phase \cite{ISS}.  
The typical model is an ${\cal N}=1$ SUSY $SU(N_c)$ gauge theory 
with massive (but light) $N_f$ quark chiral multiplets 
in the range $N_c < N_f < \frac{3}{2}N_c$ \cite{Seiberg}. 
In spite that the SUSY breaking vacuum is metastable, 
their finding is quite remarkable in that it opened a new avenue 
to overcome the two major difficulties mentioned in the previous paragraph. 
First, the model building becomes very easy 
since the ISS models are vector-like. 
Second, we can guess that the Landau pole problem is relaxed 
since the ISS models has a large flavor number comparing 
to the number of color, which reduces the number of messengers. 
Motivated by the proposal of ISS, 
extensive researches on the metastable SUSY breaking have been carried out 
in various viewpoints \cite{Forste}-\cite{AKO}.

There is, however, a serious problem in the ISS model. 
$U(1)_R$ symmetry is unbroken at the metastable SUSY breaking vacuum, 
which implies that the gauginos cannot obtain a Majorana mass. 
How the $U(1)_R$ is broken is the first step to the model building 
of metastable SUSY breaking. 
Some models of direct mediation of metastable SUSY breaking 
with broken $U(1)_R$ symmetry have been proposed 
using the inverted hierarchy mechanism \cite{DM, CST} and 
the explicit $U(1)_R$ breaking superpotential term \cite{KOO, ADJK}.

In this paper, we construct a direct gauge mediation model of metastable 
SUSY breaking by simply deforming the ISS model 
in terms of a dual meson superpotential mass term. 
No extra matter is introduced. 
The deformation explicitly breaks $U(1)_R$ symmetry. 
We show that the vacuum expectation value (VEV) 
of the pseudo moduli is located at the nonzero value and 
messenger fields obtain  SUSY masses through the Coleman-Weinberg potential. 
Our metastable SUSY breaking vacuum turns out to be sufficiently long-lived. 
Gauging a subgroup of the flavor symmetry and 
identifying it with the SM gauge group, 
the direct gauge mediation is realized and 
nonvanishing gaugino masses are generated. 
We also find that Landau pole problem can be avoided. 
These analyses show 
the parameters in the dynamical SUSY breaking sector are
phenomenologically constrained. 

%%%%%%%%%%%%%%%
\section{Model}
%%%%%%%%%%%%%%%
In this section, 
we consider a metastable SUSY breaking model 
of ISS deformed by a dual meson superpotential mass term 
which breaks a $U(1)_R$ explicitly. 
The model is an ${\cal N}=1$ SUSY $SU(N_f-N_c)$ gauge theory 
with $N_f$ flavor dual quarks $q$, $\tilde{q}$ 
and the gauge singlet $M$ couples to the dual quarks 
in the superpotential, which is a free magnetic description of 
SUSY $SU(N_c)$ gauge theory with massive $N_f$ flavors. 
The superpotential and K\"ahler potential are given by
\bea
W &=& {\rm Tr}(q M \tilde{q}) + m^2 {\rm Tr}M 
+ \frac{\epsilon}{\Lambda} {\rm Tr}(q \tilde{q})^2, 
\label{spot} \\
K &=& M^\dag M  + q^\dag q + \tilde{q} \tilde{q}^\dag, 
\label{kahler}
\eea
where $m$ and $\Lambda$ are dimensionful parameters 
and $\epsilon$ is a dimensionless parameter. 
The ISS model is just deformed by the last term 
in the superpotential (\ref{spot}), 
which explicitly breaks the $U(1)_R$ symmetry. 
Note that the simplest deformation by the term ${\rm Tr}q \tilde{q}$ 
does not change the physics in the ISS model, 
namely the $U(1)_R$ is unbroken even at one loop level 
since this term can be absorbed into the first term in (\ref{spot}) 
by the constant shift of $M$.

Parametrizing $M$, $q$, $\tilde{q}$ and $m$ as, 
\bea
&&M = 
\left(
\begin{array}{cc}
Y & Z \\
\tilde{Z} & \Phi \\
\end{array}
\right), \quad 
q = 
\left(
\begin{array}{c}
\chi \\
\rho \\
\end{array}
\right), \quad 
\tilde{q} = 
\left(
\begin{array}{c}
\tilde{\chi} \\
\tilde{\rho} \\
\end{array}
\right), \quad 
m = 
\left(
\begin{array}{cc}
m \delta_{ab} & 0 \\
0 & \tilde{m} \delta_{AB} \\
\end{array}
\right) 
\eea
where $a,b = 1, \cdots, N_f-N_c$ and $A,B=1,\cdots,N_c$ 
and we assumed $m$ to be in the block diagonal form. 
%${\bf 1}_{N, N}$ denotes the $N \times N$ unit matrix. 
The superpotential and K\"ahler potential are decomposed as 
\bea
W &=& {\rm Tr} 
\left[ 
\chi Y \tilde{\chi} +\chi Z \tilde{\rho} 
+ \tilde{\chi} \tilde{Z} \rho + \rho \Phi \tilde{\rho} 
+ m^2 Y + \tilde{m}^2 \Phi \right] \nonumber \\
&&+\frac{\epsilon}{\Lambda}
{\rm Tr}\left[
(\chi \tilde{\chi})^2 + (\chi \tilde{\rho})(\tilde{\chi}\rho) 
+ (\chi \tilde{\chi})(\chi \tilde{\rho}) 
+ (\chi \tilde{\rho})(\rho \tilde{\rho}) 
+ (\rho \tilde{\chi})(\chi\tilde{\chi}) \right. \nonumber \\
&& \left. + (\rho \tilde{\rho})(\rho \tilde{\chi}) 
+ (\rho\tilde{\chi})(\chi\tilde{\rho}) + (\rho \tilde{\rho})^2
\right], 
\label{spotdeco}\\
K &=& |Y|^2 + |Z|^2 + |\tilde{Z}|^2 + |\Phi|^2 
+ |\chi|^2 + |\rho|^2 + |\tilde{\chi}|^2 + |\tilde{\rho}|^2. 
\label{kahlerdeco}
\eea
Let us study the vacuum structure at the classical level. 
Using the degree of freedom of the symmetries 
$SU(N) \times SU(N_f) \times SU(N_f)$, 
we can take the VEVs of the matter fields 
in the following block diagonal form, 
namely $Z=\tilde{Z} = \rho = \tilde{\rho} = 0$, and 
\bea
M = 
\left(
\begin{array}{cc}
Y_{ab}\delta_{ab} & 0 \\
0 & \Phi_{AB}\delta_{AB} \\
\end{array}
\right), \quad 
q = 
\left(
\begin{array}{c}
\chi_{ab}\delta_{ab} \\
0 \\
\end{array}
\right), \quad 
\tilde{q} = 
\left(
\begin{array}{c}
\tilde{\chi}_{ab}\delta_{ab} \\
0 \\
\end{array}
\right). 
\label{classicalms}
\eea
%where $a,b = 1, \cdots, N_f-N_c$ and $A,B=1,\cdots,N_c$. 
Note that the gauge symmetry $SU(N_f-N_c)$ is completely broken in this vacuum. 
Putting these VEVs into F-flatness conditions leads to 
\bea
0 &=& \frac{\partial W}{\partial Y} 
= \chi \tilde{\chi} + m^2, 
\label{FY1}\\ 
0 &\ne& \frac{\partial W}{\partial \Phi} = \tilde{m}^2, 
\label{FPhi1} \\ 
0 &=& \frac{\partial W}{\partial \chi} 
= \left[ Y + \frac{2\epsilon}{\Lambda} (\chi \tilde{\chi}) \right] 
\tilde{\chi}, 
\label{Fchi1}\\ 
0 &=& \frac{\partial W}{\partial \rho} 
= \frac{\epsilon}{\Lambda} \tilde{\chi} (\chi \tilde{\chi}), 
\label{Frho1}\\ 
0 &=& \frac{\partial W}{\partial \tilde{\chi}} 
= \left[ Y + \frac{2\epsilon}{\Lambda} (\chi \tilde{\chi}) \right] \chi, 
\label{Ftildechi1}\\ 
0 &=& \frac{\partial W}{\partial \tilde{\rho}} 
= \frac{\epsilon}{\Lambda} \chi (\chi \tilde{\chi}). 
\label{Ftilderho1}
\eea
The second equation $F_\Phi \ne 0$ shows that SUSY is spontaneously broken. 
Its fermionic component $\psi_\Phi$ is a Nambu-Goldstone fermion and 
eaten by the gravitino when the gravitational coupling is switched on. 

The scalar potential at tree level is given by 
\bea
V_{{\rm tree}} = |\chi \tilde{\chi} + m^2|^2 + |\tilde{m}^2|^2 
+ \left[ 
\left| Y + \frac{2\epsilon}{\Lambda} \chi \tilde{\chi} \right|^2 
+ \left| \frac{\epsilon}{\Lambda}(\chi \tilde{\chi}) \right|^2 
\right]
(|\chi|^2 + |\tilde{\chi}|^2). 
\label{potential}
\eea
This potential has a local minimum with
\bea
Y + \frac{2\epsilon}{\Lambda}\chi \tilde{\chi} = 0
\label{case1}
\eea
and a local maximum with 
\bea
|\chi| = |\tilde{\chi}| = 0,~Y~({\rm arbitrary})
\label{case2}
\eea
where $Y$ is undetermined at tree level. 
From now, we focus on the case (\ref{case1}). 
The minimization conditions in this case is given by 
\bea
0 &=& \chi \tilde{\chi} + m^2 
+ 2\left(\frac{\epsilon}{\Lambda} \right)^2 \chi \tilde{\chi} |\chi|^2 
+ \left|\frac{\epsilon}{\Lambda} \chi \tilde{\chi} \right|^2 
\label{minicase1}
\eea
where D-flat condition $|\chi|=|\tilde{\chi}|$ is used. 
Expanding VEVs as 
%\footnote{ISS case is recovered in the limit $\epsilon \to 0$.}
\bea
\chi \tilde{\chi} &=& - m^2 + \epsilon (\chi \tilde{\chi})_1 
+ \epsilon^2 (\chi \tilde{\chi})_2 +{\cal O}(\epsilon^3), \\
Y &=& \frac{2\epsilon}{\Lambda}m^2 + {\cal O}(\epsilon^3), 
\eea
where $(\chi \tilde{\chi})_{1,2}$ denotes 
${\cal O}(\epsilon^{1,2})$ part of $\chi \tilde{\chi}$, 
we obtain
\bea
&& (\chi\tilde{\chi})_1 = 0, \\
&& (\chi \tilde{\chi})_2 = \frac{m^2}{\Lambda^2}(2|\chi|^2 - m^2) 
\approx \frac{m^4}{\Lambda^2}. 
\eea
The vacuum energy $V_{{\rm local~minimum}}$ at the local minimum (\ref{case1}) 
and $V_{{\rm local~maximum}}$ at the local maximum (\ref{case2}) 
can be calculated as 
\bea
V_{{\rm local~minimum}} &=& 
(N_f - N_c)|\langle {\cal O}(\epsilon^2) \rangle|^2 
+ N_c|\tilde{m}^2|^2 + (N_f-N_c)\frac{2\epsilon^2}{\Lambda^2}m^6, \\
V_{{\rm local~maximum}} &=& (N_f-N_c)|m^2|^2 + N_c|\tilde{m}^2|^2,  
\eea
which will be necessary to compute the lifetime of 
our metastable SUSY breaking vacuum.

%%%%%%%%%%%%%%%%%%%%%%%%%%%%%%%%%%%%%%%%%%%%%%%%%%%%%%%%%%%%%%%%%%%%%%%%%%%%%%%%
\section{Calculation of Coleman-Weinberg potential for the pseudo-moduli $\Phi$}
%%%%%%%%%%%%%%%%%%%%%%%%%%%%%%%%%%%%%%%%%%%%%%%%%%%%%%%%%%%%%%%%%%%%%%%%%%%%%%%%
At tree level, $\langle \Phi \rangle$ is undetermined 
since $\Phi$ does not appear in the tree level potential. 
Because of $F_\Phi \ne 0$, if the $\langle \Phi \rangle$ is lifted 
at the finite value by one-loop quantum corrections, 
SUSY breaking can be transmitted to the MSSM sector. 
%What is crucial is to investigate whether this situation is realized 
%in our model. 
To study the issue, we have to compute a Coleman-Weinberg potential 
for the moduli $\Phi$. 
All the massive fields at tree level contribute to the potential. 
Since we are interested in the moduli $\Phi$, 
we focus on the massive fields, $\rho, \tilde{\rho}, Z, \tilde{Z}$, 
whose masses depend on the VEV of $\Phi$. 

The mass matrix for their bosonic components is calculated as 
\bea
(\rho^\dag, \tilde{\rho}^\dag, Z^\dag, \tilde{Z}^\dag){\cal M}_B^2 
\left(
\begin{array}{c}
\rho \\
\tilde{\rho} \\
Z \\
\tilde{Z} \\
\end{array}
\right), \quad 
{\cal M}_B^2 \equiv 
\left(
\begin{array}{cc}
W^{\dag i k}W_{kj} & W^{\dag ijk}W_k \\
W_{ijk}W^{\dag k} & W_{i k}W^{\dag k j} \\
\end{array}
\right), 
\eea
where 
\bea
W^{\dag i k}W_{kj} &\approx& 
\left(
\begin{array}{cccc}
m^2 + |\Phi|^2 & 4 \left( \frac{\epsilon}{\Lambda} \right)^2 m^4 & m \Phi & 0 \\
4 \left( \frac{\epsilon}{\Lambda} \right)^2 m^4 & m^2 + |\Phi|^2 & 0 & m \Phi \\
m \Phi^\dag & 0 & m^2 & 0 \\
0 & m \Phi^\dag & 0 & m^2 \\
\end{array}
\right), \\
W^{\dag i jk}W_{k} &\approx& 
\left(
\begin{array}{cccc}
2 \left( \frac{\epsilon}{\Lambda} \right)^2 m^4 & \tilde{m}^2 & 0 & m Y \\
\tilde{m}^2 & 2 \left( \frac{\epsilon}{\Lambda} \right)^2 m^4 & mY & 0 \\
0 & mY^\dag & 0 & 0 \\
mY^\dag & 0 & 0 & 0 \\
\end{array}
\right) 
\eea
where $W_a \equiv \partial W/\partial \phi^a$ etc. 
The corresponding mass matrix for fermionic components ${\cal M}_F$ 
are obtained 
by taking the zero limit of SUSY breaking $\tilde{m}^2=0$. 
The parameter range without tachyonic messenger mass squareds 
is given as follows. 
\bea
%4 \epsilon^2 \left(\frac{m}{\Lambda} \right)^2 
-1 + \frac{\tilde{m}^2}{m^2} 
\pm {\cal O}(1) \epsilon^2 \left(\frac{m}{\Lambda} \right)^2  
< 4\epsilon \frac{m}{\Lambda} \frac{\Phi}{m} < 
%-4 \epsilon^2 \left(\frac{m}{\Lambda} \right)^2 
+ 1 + \frac{\tilde{m}^2}{m^2} \pm {\cal O}(1) 
\epsilon^2 \left(\frac{m}{\Lambda} \right)^2. 
\eea
We search for a local minimum with nonvanishing VEV of 
the pseudo moduli $\Phi$ in this range of parameters. 

The Coleman-Weinberg potential is defined as 
\bea
V_{{\rm 1-loop}} = \frac{1}{64\pi^2} {\rm Tr}
\left[ {\cal M}_B^4 \ln \frac{{\cal M}_B^2}{\Lambda^2} 
- {\cal M}_F^4 \ln \frac{{\cal M}_F^2}{\Lambda^2} \right]
\label{CW}
\eea
where $\Lambda$ is a cutoff scale. 
All massive fields except for a pseudo-moduli $\Phi$ contribute to 
the Coleman-Weinberg potential. 
$Y, \chi, \tilde{\chi}$ have ${\cal O}(m)$ masses. 
$\rho, \tilde{\rho}, Z, \tilde{Z}$ have masses depending on 
the $\langle \Phi \rangle$ and the masses are turned out to be 
$m_\pm \sim {\cal O}(m)~{\rm or}~{\cal O}(0.1m)$ discussed later. 
We found that the minimum is shifted from the origin. 
One example is shown in Fig. \ref{potential} of section 6. 
In this numerical computation, we took all parameters to be real. 
As a consistency check, 
taking the limit $\epsilon=0, \tilde{m}^2=m^2$, 
we confirmed that the ISS results $\langle \Phi \rangle = 0$ is reproduced.

%%%%%%%%%%%%%%%%%%%%%%%%%%%%%%%%%%%%%%%%%%%%%%%
\section{Mediation of Metastable SUSY breaking}
%%%%%%%%%%%%%%%%%%%%%%%%%%%%%%%%%%%%%%%%%%%%%%%
In the previous section, we found $\langle \Phi \rangle \ne 0$ at 1-loop, 
which implies that the gaugino mass can be generated 
if $\rho, \tilde{\rho}$ are identified with messengers and the standard model 
gauge group is embedded in the unbroken subgroup of the flavor symmetry 
$SU(N_c)$. 

In our case, the mass matrix for the messengers takes of the form
\bea
W \supset (\rho, Z)
\left(
\begin{array}{cc}
\Phi - 2\frac{\epsilon}{\Lambda}m^2 & m \\
m & 0 \\
\end{array}
\right)
\left(
\begin{array}{c}
\tilde{\rho} \\
\tilde{Z} \\
\end{array}
\right) 
\equiv 
(\rho, Z)
{\cal M}
\left(
\begin{array}{c}
\tilde{\rho} \\
\tilde{Z} \\
\end{array}
\right). 
\label{messenger}
\eea
The leading gaugino masses at the linear order of SUSY breaking $F_\Phi$ 
are given by the formula 
$m_{\lambda_i} = \frac{g_i^2}{(4\pi)^2}N F_\Phi \frac{\partial}{\partial \Phi}
{\rm log}{\rm det}{\cal M}$ 
where $N$ is a flavor number of messengers. 
Note that this leading gaugino masses vanish because 
${\rm det}{\cal M}$ does not depend on $\Phi$ in (\ref{messenger}). 

Then, the gaugino masses in our case are generated 
at one-loop by the cubic order of SUSY breaking 
${\cal O}(F^3_\Phi)$ \cite{INTY}. 
On the other hand, the sfermion masses are generated 
in a usual gauge mediation form: 
\bea
&&m_{\lambda_i} \approx (N_f - N_c) \frac{\alpha_i}{4\pi} 
\left(\frac{\langle F_\Phi \rangle}{\langle \Phi \rangle^2} \right)^2 
\frac{\langle F_\Phi \rangle}{\langle \Phi \rangle} 
\sim (N_f - N_c) \frac{\alpha_i}{4\pi} 
\left(\frac{\tilde{m}}{m} \right)^6 m, 
\label{gaugino} \\ 
&&m^2_{\tilde{f}} \approx (N_f - N_c) C_i \left( \frac{\alpha_i}{4\pi} \right)^2 
\left(\frac{\langle F_\Phi \rangle}{\langle \Phi \rangle} \right)^2 
\sim 
(N_f - N_c) C_i \left( \frac{\alpha_i}{4\pi} \right)^2 
\left(\frac{\tilde{m}}{m} \right)^4 m^2, 
\label{sfermion}
\eea
where $\alpha_i$ are fine structure constants 
for the standard model gauge group 
$i = SU(3)_C$, $SU(2)_L$, $U(1)_Y$. 
$C_i$ are the corresponding quadratic Casimir coefficients. 
We can see that the sfermion masses are heavier than the gaugino masses, 
\bea
\frac{m_{\tilde{f}}}{m_\lambda} \sim \sqrt{\frac{C_i}{N_f-N_c}}
\left(\frac{m}{\tilde{m}} \right)^4. 
\eea
To avoid a huge hierarchy between the gaugino masses and the sfermion ones, 
the ratio $\tilde{m}^2/m^2$ must be at most 0.1. 
Otherwise, the large one-loop corrections to Higgs mass 
by the third generation squarks require unnatural tuning of parameters 
to solve the gauge hierarchy problem. 
To obtain $m_\lambda \sim {\cal O}(100~{\rm GeV})$, we need 
\bea
\left(\frac{\tilde{m}}{m} \right)^6 m \sim 10~{\rm TeV}. 
%\Rightarrow m = 10^4~{\rm TeV}~{\rm for}~\tilde{m}^2/m^2 = 0.1. 
\eea
For instance, 
the sfermion masses become 10 TeV for $\tilde{m}^2/m^2 \sim 0.1$.

The gravitino mass, which is the lightest superparticle 
in the gauge mediation, is estimated as 
\bea
m_{3/2} \simeq \frac{F_\Phi}{\sqrt{3}M_P} 
\sim 10^{-11} 
\left(\frac{m}{\tilde{m}} \right)^{10}~{\rm GeV} 
\eea
where $M_P$ is a reduced Planck scale $2.4 \times 10^{18}$ GeV. 
For instance, the gravitino mass becomes $m_{3/2} \sim 1$ keV 
for $\tilde{m}^2/m^2 \sim 0.1$, which avoids the gravitino problem 
\cite{MMY}. 

%%%%%%%%%%%%%%%%%%%%%%%%%%%%%%%%%%%%%%%%%%%%%%%%
\section{The lifetime of the Metastable vacuum}
%%%%%%%%%%%%%%%%%%%%%%%%%%%%%%%%%%%%%%%%%%%%%%%%
We have to check whether our SUSY breaking vacuum is long-lived 
compared to the age of universe 
since our SUSY breaking vacuum is not a global minimum of the potential 
but a local one. 
To estimate the decay rate from our false vacuum to the true SUSY vacuum, 
we need the VEV of a SUSY vacuum. 

Let us consider the case $\langle M \rangle \ne 0$ 
where the dual quarks $q, \tilde{q}$ are decoupled 
in the low energy effective theory below the scale $\langle M \rangle$. 
In other words, the low energy effective theory is 
an $SU(N_f-N_c)$ super Yang-Mills theory. 

SUSY conditions for $q, \tilde{q}$ are
\bea
0 &=& \frac{\partial W}{\partial q} 
= \left(M + 2 \frac{\epsilon}{\Lambda} q\tilde{q} \right)\tilde{q}, \\
0 &=& \frac{\partial W}{\partial \tilde{q}} 
= q \left(M + 2 \frac{\epsilon}{\Lambda} q\tilde{q} \right). 
\eea
We find two vacua. 
\bea
q=\tilde{q}=0~({\rm vacuum~(i)}), \quad 
M + 2\frac{\epsilon}{\Lambda}q \tilde{q} = 0~({\rm vacuum~(ii)}). 
\eea
In the vacuum (i) case, the effective superpotential becomes
\bea
W_{{\rm eff}} = m^2{\rm Tr}M + (N_f-N_c)\Lambda_L^3 
= m^2{\rm Tr}M 
+ (N_f-N_c)[\Lambda^{3(N_f-N_c)-N_f}{\rm det}M]^{\frac{1}{N_f-N_c}}
\label{weff1}
\eea
where $\Lambda_L$ is a dynamical scale of $SU(N_f-N_c)$ super Yang-Mills theory 
and the second term is generated 
by the gaugino condensation of super Yang-Mills theory. 
1-loop matching condition for the holomorphic gauge coupling 
at the scale $\langle M \rangle$, 
\bea
\Lambda_L^{3(N_f-N_c)} = \Lambda^{3(N_f-N_c)-N_f}{\rm det}M
\eea
is used in the second equality. 
SUSY condition for $M$ 
\bea
0=\frac{\partial W_{{\rm eff}}}{\partial M} 
= m^2 + \Lambda^{\frac{3(N_f-N_c)-N_f}{N_f-N_c}}
({\rm det}M)^{\frac{1}{N_f-N_c}}/M
\eea
tells us that the SUSY vacuum is located at 
\bea
\langle M \rangle \sim 
\left(\frac{m}{\Lambda} \right)^{\frac{2(N_f-N_c)}{N_c}}\Lambda.
\label{SUSYvac1}
\eea
In the vacuum (ii) case, the effective superpotential takes the form 
\bea
W_{{\rm eff}} &=& -\frac{\Lambda}{4\epsilon}{\rm Tr}M^2 
+ m^2 {\rm Tr}M + (N_f-N_c)\Lambda_L^3 \nonumber \\
&=& -\frac{\Lambda}{4\epsilon}{\rm Tr}M^2 
+ m^2 {\rm Tr}M 
+ (N_f-N_c)[\Lambda^{3(N_f-N_c)-N_f}{\rm det}M]^{\frac{1}{N_f-N_c}}. 
\eea
SUSY condition for $M$ 
\bea
0=\frac{\partial W_{{\rm eff}}}{\partial M} 
= -\frac{\Lambda}{2\epsilon}M + m^2 
+ \Lambda^{\frac{3(N_f-N_c)-N_f}{N_f-N_c}}
M^{\frac{N_c}{N_f-N_c}}
\eea
provides
\bea
\langle M \rangle \sim 
\left(\frac{1}{2\epsilon}\right)^{\frac{N_f-N_c}{2N_c-N_f}} \Lambda
\label{SUSYvac2}
\eea
where $m^2$ in the SUSY condition is neglected. 
%Interestingly, $\langle M \rangle$ depends on $\epsilon$ in this case 
%and goes to infinity in the ISS limit. 

We are now in a position to estimate the decay rate 
from the false vacuum to the true vacuum. 
The necessary ingredients to estimate are the potential height $V_{{\rm peak}}$ 
and the distance between the false vacuum and the true one 
in the field space $\Delta \Phi$. 
In our case, the potential height at the corresponding local maximum 
\bea
\langle \Phi \rangle \simeq m, \quad 
\langle Y \rangle = \frac{2\epsilon}{\Lambda} m^2, \quad 
|\chi| = |\tilde{\chi}| = 0
\eea
is 
\bea
V_{{\rm peak}} = (N_f-N_c)|m^2|^2 + N_c |\tilde{m}^2|^2 \approx |m^2|^2. 
\eea
The field distance between two vacua is 
\bea
\Delta \Phi 
%&=& \langle M \rangle_{{\rm SUSY~vacuum}} 
%- \langle \Phi \rangle_{{\rm SUSY~breaking}} \nonumber \\
\approx \langle M \rangle_{{\rm SUSY~vacuum}} 
= \left\{
\begin{array}{c}
\left( \frac{m}{\Lambda} \right)^{\frac{2(N_f-N_c)}{N_c}} 
\Lambda~({\rm vacuum~(i)}), \\
\left(\frac{1}{2\epsilon} \right)^{\frac{N_f-N_c}{2N_c-N_f}} 
\Lambda~({\rm  vacuum~(ii)}). \\
\end{array}
\right. 
\eea
Then, the bounce action is estimated as
\bea
S \sim \frac{(\Delta \Phi)^4}{V_{{\rm peak}}} 
\sim 
\left\{
\begin{array}{c}
\left(
\frac{\Lambda}{m} \right)^{4-8(N_f - N_c)/N_c} \gg 1~({\rm vacuum~(i)}), \\
\left(\frac{1}{2\epsilon} \right)^{4(N_f-N_c)/(N_c-(N_f-N_c))} 
\left( \frac{\Lambda}{m} \right)^4 \gg 1~({\rm  vacuum~(ii)}). \\
\end{array}
\right. 
\eea
The bounce action can be parametrically large 
as long as we take $m/\Lambda \ll 1$. 

For instance, let us consider the case $m/\Lambda = 0.1, \epsilon=1, 
N_f =6$ and $N_c = 5$ which gives the most stringent bound, 
then we obtain 
\bea
S > 
%\left(\frac{\Lambda}{m} \right)^{4(N_c-2)/3} > 
10^{2.4}~({\rm vacuum~(i)}), \quad
%\left( \frac{1}{2\epsilon} \right)^{4/(N_c-2)} 
%\left(\frac{\Lambda}{m} \right)^4 > 
4 \times 10^3~({\rm vacuum~(ii)}). 
\eea
%where $N_c \ge 5$ is also taken into account. 
The lifetime of the metastable vacuum can be obtained 
\bea
\tau \sim e^S > e^{251}~({\rm vacuum~(i)}),\quad 
e^{4000}~({\rm vacuum~(ii)}) 
\eea
which implies that our SUSY breaking vacuum is sufficiently long-lived 
compared to the age of the universe ($\sim e^{40}$). 

%%%%%%%%%%%%%%%%%%%%%%%%%%%%%%
\section{Landau pole analysis}
%%%%%%%%%%%%%%%%%%%%%%%%%%%%%%
In this section, we examine whether the QCD coupling constant 
at the GUT scale is perturbative or not when the standard model gauge group 
is embedded in the unbroken subgroup of the flavor symmetry $SU(N_c)$.

One-loop gauge coupling RGE is given by 
\bea
g_i^{-2}(\mu) = g_i^{-2}(\mu') + \frac{b_i}{8\pi^2}
\ln \left(\frac{\mu}{\mu'} \right)
\eea
where $b_i$ is one-loop beta function coefficient of the gauge group 
$i=SU(3)_c, SU(2)_L, U(1)_Y$. 

The one-loop beta function coefficients for QCD coupling 
at various scales are listed below. 
\bea
\mu < m_\lambda &:& b_3 = b_3^{{\rm SM}} = 7 \nonumber \\
m_{\lambda} < \mu < m_{\tilde{f}} \sim (m/\tilde{m})^4 m_\lambda 
&:& b_3 = b_3^{{\rm SM}} - \frac{2}{3} \times 3 = 5 \nonumber \\
m_{\tilde{f}} \sim (m/\tilde{m})^4 m_\lambda 
< \mu < m_- &:& b_3 = b^{{\rm MSSM}} = 3 \nonumber \\
m_- < \mu < m_+ &:& b_3 = b_3^{{\rm MSSM}} - b_3^\Phi - (N_f - N_c) 
= - (N_f - N_c) \nonumber \\
m_+ < \mu < \Lambda &:& b_3 = -2(N_f - N_c) 
\eea
where $b^{{\rm SM}}$, $b^{{\rm MSSM}}$ are the QCD one-loop beta function 
coefficients for the standard model and the minimal SUSY standard model. 
$\pm$ of $m_{\pm}$ means that the sign of the square root 
in the mass eigenvalues for messengers. 
Numerically, it is roughly $m_+ \sim m, m_- \sim 0.1m$. 
$\Phi$ is an adjoint representation under $SU(N_c)$, 
which gives $b_3^\Phi=3$ for QCD color gauge group $SU(3)_c$. 
Here we use the numerically obtained mass $m_\Phi \sim 0.1m$. 

Taking the following values 
\bea
%\mu \sim M_Z, \quad 
\Lambda = M_{{\rm GUT}} \sim 10^{16}~{{\rm GeV}}, \quad 
\frac{g_3^2(M_Z)}{4\pi} \sim 0.18, 
\eea
and one-loop beta function coefficients obtained above, 
the QCD coupling is expressed as 
\bea
\alpha_3(M_{{\rm GUT}})^{-1} 
&\sim& 
8.5 - \frac{1}{2\pi}
\left[
-0.74 
+\left\{ 13 + 6(N_f-N_c) \right\} \ln \left(\frac{\tilde{m}}{m} \right)^2 
\right. \nonumber \\
&& \left. + 2.3 \left\{ 25(N_f - N_c) - 3 \right\} 
\right]. 
\eea
Requiring that the QCD coupling constant is perturbative at the GUT scale, 
the constraint for the flavor number of messengers $N_f-N_c$ 
and $\tilde{m}^2/m^2$ can be found, 
\bea
\alpha_3(M_{{\rm GUT}}) < 1 
\Leftrightarrow \left[13 + 6(N_f - N_c) \right] 
\ln \left(\frac{\tilde{m}}{m} \right)^2 + 57.5(N_f - N_c) < 61.1. 
\label{constraint}
\eea
The examples satisfying (\ref{constraint}) are 
$N_f - N_c \le 2$ for $(\tilde{m}/m)^2=0.1$ and 
$N_f-N_c \le 4$ for $(\tilde{m}/m)^2 = 0.01$. 
The smaller $(\tilde{m}/m)^2$ allows a larger number of messengers $N_f-N_c$, 
but the hierarchy between the gaugino masses and the sfermion ones 
becomes larger. 
Thus, 
\bea
N_f-N_c = 1,2
\eea 
are phenomenologically favorable. 
It is important to mention the differences between our model and 
the other relevant models that 
Landau pole problem is avoided when the embedding of 
the standard model gauge group in $SU(N_f-N_c)$ not in $SU(N_c)$ 
in the model \cite{KOO} and cannot be avoided in the model \cite{ADJK}. 
Taking into account the facts that the free magnetic description is valid 
for $N_c+1 \le N_f < \frac{3}{2}N_c$ and $SU(N_c)$ embedding 
of the standard model gauge group is possible only for $N_c \ge 5$, 
the allowed ranges of the number of colors and flavors are found
\bea
5 \le N_c, \quad 6 \le N_f < \frac{3}{2}N_c. 
\eea

A numerical result for a metastable SUSY breaking minimum 
found in this analysis is summarized 
in the table and Fig.~\ref{potential} below. 
\bea
\begin{array}{|c|c|c|}
\hline
\langle \Phi \rangle/m & -1.18 & -1.18 \\
\hline
\tilde{m}^2/m^2 & 0.1 & 0.1 \\
\hline
\epsilon & 1 & 1 \\
\hline
m/\Lambda & 0.1 & 0.1 \\
\hline
m_\lambda~({\rm GeV}) & 100 & 100 \\
\hline
m_{\tilde{f}}~({\rm TeV}) & 10 & 10 \\
\hline
{\rm lifetime} & \gg \tau_0 & > \tau_0(N_c \ge 8) \\
\hline
N_f-N_c & 1 & 2 \\
\hline
\end{array}
\nonumber 
\eea
Here $\tau_0$ means the age of universe. 
%%%%%%%%%%%%%%%%%
\begin{figure}[htb]
 \begin{center}
  \includegraphics[width=5.5cm]{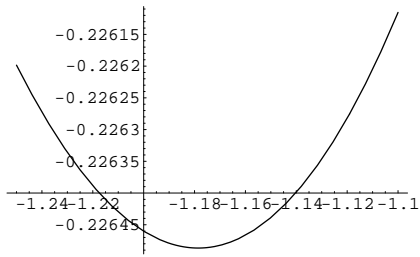}
%\hspace*{10mm}
%  \includegraphics[width=5cm]{potential2.eps}
 \end{center}
 \vspace*{-0.5cm}
\caption{One-loop Coleman-Weinberg potential for the pseudo moduli $\Phi$. 
The vertical axis denotes $V_{{\rm 1-loop}}/m^4$, 
and the horizontal axis does $\Phi/m$. 
%These plots correspond to the cases 1,2 from the left to the right.
} 
\label{potential}
\end{figure}
%%%%%%%%%%%%%%%%%%%%
The constraint of color number in $N_f - N_c = 2$ case 
comes from 
the lifetime of our metastable SUSY breaking vacuum 
(decay to the vacuum (i)) and 
the allowed region from Landau pole analysis,
\bea
\left(\frac{\Lambda}{m} \right)^{4-8(N_f-N_c)/N_c} > 40. 
\eea
No constraint for the number of color arises from the decay to the vacuum (ii).

%%%%%%%%%%%%%%%%%%%%%
\section{Summary}
%%%%%%%%%%%%%%%%%%%%%
We have constructed a model of direct gauge mediation of 
metastable SUSY breaking by simply deforming the ISS model 
in terms of a dual meson superpotential mass term, 
which breaks an $U(1)_R$ explicitly. 
No extra matter field is introduced. 
It was shown that the VEV of the pseudo moduli $\Phi$ is shifted 
at the finite value 
through the Coleman-Weinberg potential for the pseudo moduli. 
The lifetime of our metastable vacuum was turned out to 
be sufficiently long compared to the age of universe 
as long as $m/\Lambda \ll 1$. 
By gauging a subgroup of the flavor symmetry $SU(N_c)$, 
the direct gauge mediation was realized. 
In our model, the sfermion masses are heavier than the gaugino masses. 
Typically, sfermion masses are 10 TeV corresponding to 
the SUSY breaking scale of order $10^6$ GeV. 
The condition for the QCD coupling to be perturbative at the GUT scale 
was derived and it is found that Landau pole problem can be avoided. 
From these analysis, the parameters in the SUSY breaking sector have been 
phenomenologically constrained.

It is deserved to pay an attention that our simple deformed ISS model 
provides a phenomenologically viable model of the direct gauge mediation 
of metastable SUSY breaking. 
Introducing only the term ${\rm Tr}(q \tilde{q})^2$ 
among the same order other operators is just an assumption, 
but might be explained from the view point of brane picture.  
We can say at least that SUSY is not restored even if more general 
higher dimensional operators 
${\rm Tr}(q \tilde{q})^n~(n:{\rm integer~more~than~2})$ 
are added to the superpotential since SUSY breaking condition 
$\partial W/\partial \Phi \ne 0$ is unchanged. 
This tells us that our conslusion obtained in this paper 
is essentially unchanged. 
Adding the operators including $M$ to the superpotential would spoil 
our conslusion since SUSY is restored. 
We hope that our model discussed in this paper will shed some insights 
for further studies on the model building of metastable SUSY breaking.

%%%%%%%%%%%%%%%%%%%%%%%%%%%%%%
\subsection*{Acknowledgments}
%%%%%%%%%%%%%%%%%%%%%%%%%%%%%
We thank C. Csaki, C. Durnford, Y. Hyakutake, J. Jaeckel, 
T. Kawano and R. Kitano for useful discussions. 
The work of the authors was supported 
in part by the Grant-in-Aid for Scientific Research 
of the Ministry of Education, Science and Culture, 
No.16540258 (N.H.), No.17740146 (N.H.) and No.18204024 (N.M.).  

%%%%%%%%%%%%%%%%%%%%%%%%%%%

\end{document}